\documentstyle[letters]{mn}
\font\japit = cmti10 at 10truept
\include{epsf}
\title
     [Self-consistent Gravitational Lens Reconstruction]
{\vglue-3.0truecm
\centerline{\japit For submission to Monthly Notices}
\vglue 2.5truecm
\noindent
Self-consistent Gravitational Lens Reconstruction
\author
     [ S. Dye \& A. N. Taylor ]
     {Simon Dye \& Andy Taylor \\
     Institute for Astronomy, 
     University of Edinburgh,
     Royal Observatory,
     Blackford Hill, 
     Edinburgh EH9 3HJ, 
     U.K.\\
	sd@roe.ac.uk, ant@roe.ac.uk}}

\newcommand{\be}{\begin{equation}}
\newcommand{\ee}{\end{equation}}
\newcommand{\ba}{\begin{eqnarray}}
\newcommand{\ea}{\end{eqnarray}}
\newcommand{\x}{\mbox{\boldmath $x$}}

\newcommand{\tb}{\mbox{\boldmath $\theta$}}

\newcommand{\kb}{\mbox{\boldmath $\kappa$}}
\newcommand{\bone}{\mbox{\boldmath $1$}}
\newcommand{\bA}{\mbox{\boldmath $A$}}
\newcommand{\bB}{\mbox{\boldmath $B$}}
\newcommand{\bG}{\mbox{\boldmath $G$}}
\newcommand{\nn}{\nonumber \\}
\def\bib{\parskip=0pt\par\noindent\hangindent\parindent
    \parskip =2ex plus .5ex minus .1ex}
\begin{document}

\maketitle

\begin{abstract}

We present a new method for directly 
determining accurate, self-consistent cluster 
lens mass and shear maps in the strong lensing regime from the magnification 
bias of background galaxies. The method relies upon pixellisation
of the surface mass density distribution which allows us to write down 
a simple, solvable set of equations. We also show how pixellisation can 
be applied to methods of mass determination from measurements of shear
and present a simplified method of application.
The method is demonstrated with cluster models and applied to
magnification data from the lensing cluster Abell 1689.

\end{abstract}


\section{Introduction}
The possibility of reconstructing cluster lens mass distributions from 
the magnification bias of background galaxies was
first suggested by Broadhurst, Taylor \& Peacock (1995) and first
demonstrated by Taylor et al. (1998, T98 hereafter). 
They showed how a direct, local measure of the lens
convergence, $\kappa=\Sigma/\Sigma_c$, where $\Sigma$ is the mass surface
density and $\Sigma_c$ is the critical surface density, could be obtained 
from knowledge of the lens magnification. In this way,
one could measure absolute
surface mass densities, thereby breaking the ``sheet-mass'' degeneracy found
in methods based on distortions of background galaxies (Tyson, Valdes \&
Wenk 1990, Kaiser \& Squires 1993, Seitz \& Schneider 1995).

van Kampen (1998) and T98 have shown
how one can extend magnification analysis
into the strong lensing regime. By making reasonable assumptions about 
$\gamma$, the lens shear, they showed that one could place quite stringent 
bounds on $\kappa$. In addition, T98 found an exact solution for
the profile of axisymmetric lenses, although not for more general 2D cases.

Inverse reconstruction methods based on maximum likelihood
(Bartelmann et al. 1996) and maximum entropy (Seitz, Schneider \&
Bartelmann 1998, Bridle et al. 1998) have gone some way in providing
a unification of both shear and magnification information. 
Until now however, no direct method using only magnification has existed.

In this letter, we show how to directly compute an accurate, self-consistent
2D distribution of $\kappa$ and $\gamma$ in the strong lensing regime
from magnification. This direct approach has the advantage
over indirect alternatives
that uncertainties can easily be determined.
The method is based on pixellisation of the
$\kappa$ distribution suggested by AbdelSalam, Saha \& Williams (1998)
who used it to estimate the mass of Abell 370 from multiple images.
We generalise the method further and also
derive a simplified solution to the problem of estimating mass from
shear, based on the approach of Kaiser \& Squires (1993). 

\section{Reconstruction of $\kappa$ and $\gamma$}
T98 showed how to estimate cluster surface mass
using the magnification measured from the distortion in 
background galaxy number counts. Here our problem is to find an accurate 
method for reconstructing the surface mass density, given the magnification
by an arbitrary lens.
The inverse magnification factor at a given position in the lens plane is
\be
\label{mageq}
	A^{-1} = |(1-\kappa)^2 - \gamma^2|,
\ee
where $\kappa$  is the lens convergence and $\gamma$ is the shear.
The shear can be decomposed into two orthogonal 
polarisation states, $\gamma_1$ and $\gamma_2$, which are related 
to the lens convergence by
\be
\gamma_1 = \frac{1}{2} \partial^{-2}(\partial^2_1-\partial^2_2) \kappa,
\quad \gamma_2 = \partial^{-2}\partial_1\partial_2 \kappa.
\ee
where $\partial_i\equiv \partial/\partial \theta_i$ and
$\partial^{-2}$ is the 2D inverse Laplacian.
The total shear is given by $\gamma^2=\gamma^2_1+\gamma^2_2$. 
One might expect that equation (\ref{mageq}) could be solved
iteratively by first 
estimating $\kappa$, using this to calculate $\gamma$ and then updating 
the estimate of $\kappa$ using equation (\ref{mageq}) again.
This proves to be highly unstable in the strong lensing regime however,
rapidly diverging after only a few iterations (Seitz \& Schneider 1995).

To find a stable solution to equation (\ref{mageq}),
we first pixellise the image.
Following AbdelSalam et al. (1998), we can now write
\be
\label{pix_gam}
	\gamma_i^n = D_i^{mn} \kappa_m, \quad  i=1,2
\ee
with summation implied over index $m$ and
where $\kappa_m$ and $\gamma_i^n$ are the pixellised
convergence and shear distributions respectively.
The transformation matrices, $D_i^{mn}$, are 
\ba
	D_1^{mn} &=& \frac{1}{2} (\partial_1^2 - \partial^2_2) 
		\int_{m}{\rm d}^2 \theta^{'}\ln |\tb_{n}-\tb^{'}|\nn
		 &=& \frac{1}{\pi} \tan^{-1} 
	\left[\frac{x_1^2-x_2^2}{(x_1^2+x_2^2)^2-1/4}\right],
\ea
and
\ba
	D_2^{mn} &=& \partial_1\partial_2 
		\int_{m}{\rm d}^2\theta^{'}\ln|\tb_{n}-\tb^{'}| \nn
		 &=&\frac{1}{2\pi} 
 	\ln\left[\frac{(x_1^2+x_2^2)^2-2x_1x_2+1/4}{
		(1/2+x_1^2+x_2^2)^2-(x_1-x_2)^2}\right],
\ea
with the integration acting over the $m^{th}$ pixel.
$\x=\tb_n-\tb_m$ is the difference between pixels $m$ and 
$n$ which are assumed to be square in calculating these
analytic expressions. Equation (\ref{mageq}) can now be written as
the vector equation,
\be
\label{matrix_eqn}
\bone-2\kb+\kb \bG \kb^{\rm t} - {\cal P}\bA^{-1} = 0
\ee
where $\bA^{-1}$ is the $N$-dimensional vector of pixellised
inverse magnification 
values, $\kb^{\rm t}$ is the transpose of the vector $\kb$
of pixellised convergence values
and $\bone$ is the vector $(1,1,1,\cdots)$. The matrix $\bG$ is the 
$N \times N\times N$ matrix 
\be
G_{pqn} = \delta_{pn}\delta_{qn} - D_1^{pn} D_1^{qn} - 
D_2^{pn} D_2^{qn}
\ee
where $\delta_{ij}$ is the Kr\"{o}necker delta, and summation
is only over indices $p$ and $q$. The parity of the 
measured inverse amplification $A^{-1}(\tb)$ 
is handled by ${\cal P}$ which flips from
being $+1$ outside regions bounded by critical lines to $-1$ 
within such regions.

The amplification equation in the form of equation (\ref{matrix_eqn}) 
is the first main result of this letter. We can now solve for $\kappa$
given a measured inverse amplification. Having solved for $\kappa$,
the corresponding shear distribution can then be calculated from 
equation (\ref{pix_gam}).

\section{Application to Cluster Models}
\label{application}

We apply the method to two types of idealised cluster
models. Starting with a predetermined cluster mass density distribution,
the corresponding shear distribution is derived using
Fourier methods (see for example,
Bartelmann \& Weiss 1994). From these,
the resulting magnification is calculated from equation
(\ref{mageq}) and then windowed
to remove boundary effects.
Using equation (\ref{matrix_eqn}), we solve for $\kappa$. 
$\gamma$ is then solved using equation(\ref{pix_gam}). A grid of 32 by 32
pixels is used in both models.

\subsection{Truncated Isothermal Sphere Model}

We first test the method with a simple truncated isothermal lens
model. The pixellated mass distribution is laid down using
$\kappa \propto (r+r_0)^{-1}$, where $r$ is the radial distance from the
centre of the sphere and $r_0$ is a constant.

\begin{figure}
\epsfxsize=8.4cm
{\hfill
\epsfbox{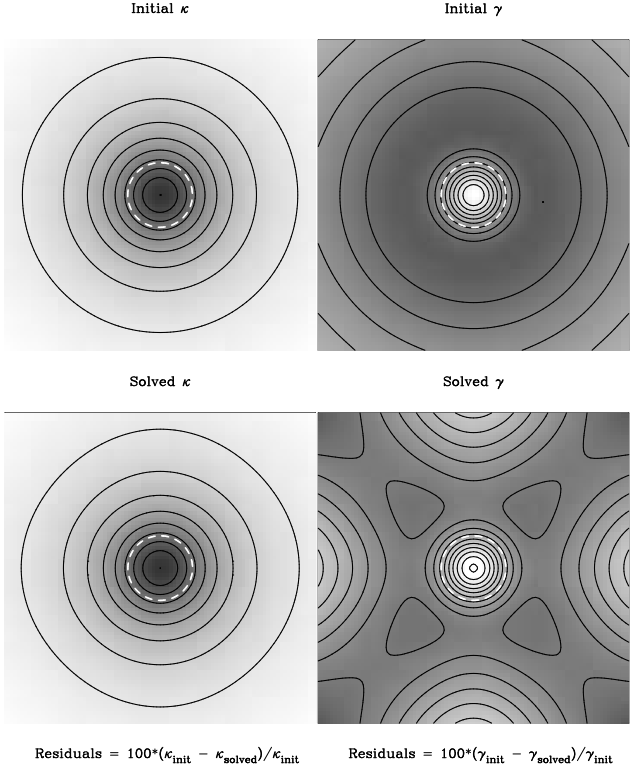}
\hfill}
\epsfverbosetrue
\small\caption{Truncated Isothermal Sphere Model: The initial $\kappa$ and
$\gamma$ used to form the magnification distribution from which the solved
$\kappa$ and $\gamma$ are derived. Underlying grid
dimensions are 32 by 32. White dashes show the
position of the critical line. Contours are linearly spaced and set at
the same levels in both $\kappa$ plots and in both $\gamma$ plots.
Residuals are expressed as percentages of
$(\kappa_{\rm init}-\kappa_{\rm solved})/\kappa_{\rm init}$}
\label{fig1}
\end{figure}

Figure \ref{fig1} shows the $\kappa$ and $\gamma$
distribution from which the magnification distribution was calculated,
the solved $\kappa$ and $\gamma$ distribution and the
difference between them.
The plotted distributions are smoothed from the underlying grid
and the white dashes highlight the lens' critical line.
The residuals are shown as percentage deviations from the true distribution. 
These are less than one percent for $\kappa$ over most of the grid 
which is negligible in comparison to the errors typically found in practice 
from background clustering, shot noise (T98) and the uncertainties
resulting from use of local $\kappa$ estimators (see van Kampen 1998).
The recovered shear distribution is more affected although
still fares better than $\gamma$ calculated from uncorrected Fourier
techniques. The main contribution to these residuals is from boundary effects
arising from trying to recover a nonlocal shear in a finite area.
Since much work has been carried out into the removal of such
effects (see Squires \& Kaiser 1996 and 
Seitz \& Schneider 1995 for example) which have little
impact on the recovered $\kappa$, we shall address the problem elsewhere.

\subsection{Dumb-bell Mass Model}

\begin{figure}
\epsfxsize=8.4cm
{\hfill
\epsfbox{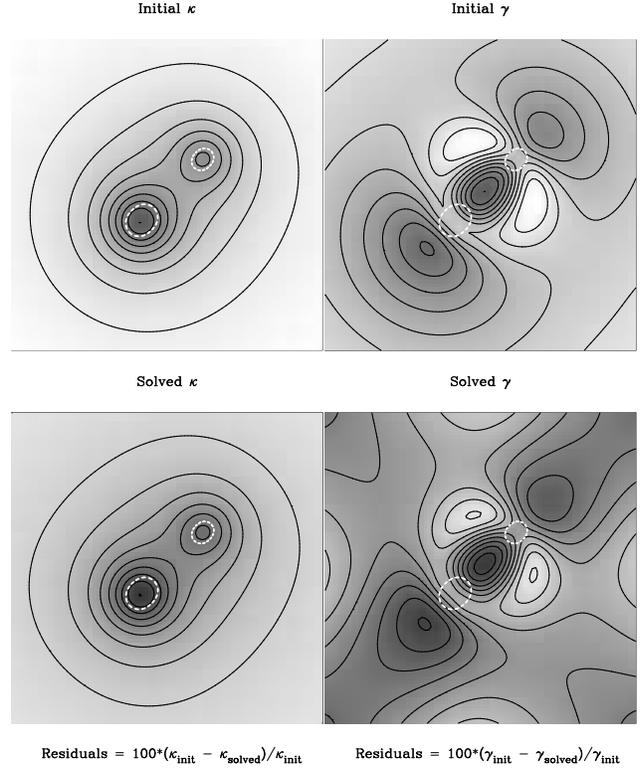}
\hfill}
\epsfverbosetrue
\small\caption{Dumb-bell Model: Critical lines are shown as white dashes.
Underlying grid dimensions are 32 by 32.
Linearly spaced contours are set at the same 
levels for $\kappa$ and at the same levels for $\gamma$.}
\label{fig2}
\end{figure}

The method was also tested with a more general dumb-bell model.
Magnification was determined in the same fashion as for the isothermal 
model, setting a negative parity
inside the critical lines, shown by the white dashes in figure \ref{fig2}.
Once again the  residuals between the
initial and solved $\kappa$ are typically less than one percent, while 
those for $\gamma$ are typically $10\%$ and again come mainly from boundary 
effects.

\section{Practical Considerations}

We solve equation (\ref{matrix_eqn}) with the hybrid
Powell method (NAG routine C05PCF). 
The number of equations needed to solve for $\kappa$ is equal to
the total number of grid pixels which can prove computationally intensive
for especially fine grids. We find that this is not
a problem for 
grid resolutions used to measure magnification
bias in practice however. The 32 by 32 grid of
pixels used for the models in section \ref{application} was solved
in approximately one minute on an average workstation. The residuals
exhibit no noticeable dependence on grid size.

The Powell algorithm is an iterative process and therefore requires an initial
estimate of the solution to start from. The choice of the initial estimate
turns out to be irrelevant. We have tried a wide range of initial
distributions and even starting from a uniform distribution arrived at the 
same final solution.

We have found that correct choice of pixel parity
(especially for low grid resolutions) is essential
in order to achieve a sensible result. Inappropriate assignment of 
parities to pixels manifest themselves, as one would expect,
by $\kappa$ being overestimated when a pixel is wrongly assumed to lie
inside a critical line and underestimated in the reverse situation.
This gives a means of checking whether critical line positions
have been properly defined by looking for large discontinuities
in the $\kappa$ distribution. Models with dual critical lines
requiring dual parity flips have also been tested and
we find that $\kappa$ can be recovered just as well.

Finally, to ensure that the method does not break down with noisy
data, we introduced a random noise term to the amplification.
Errors in $\kappa$ resulting from noise in the inverse amplification
propagate as one would expect from equation (\ref{matrix_eqn}). For 
an isothermal lens we recovered the expected result,
$\delta \kappa = \delta A/2A^2$, indicating that pixellisation does not
lead to spurious noise properties.

\section{Application to Abell 1689}

We apply the method to the magnification data presented in T98 for the 
lensing cluster A1689. A 12 by 12 grid is used as the best compromise
between shot noise in galaxy counts per bin and the resolution of the
derived $\kappa$ map. Identification of the critical line was achieved
by locating giant arc positions in the observed image.

\begin{figure}
\epsfxsize=8.4cm
{\hfill
\epsfbox{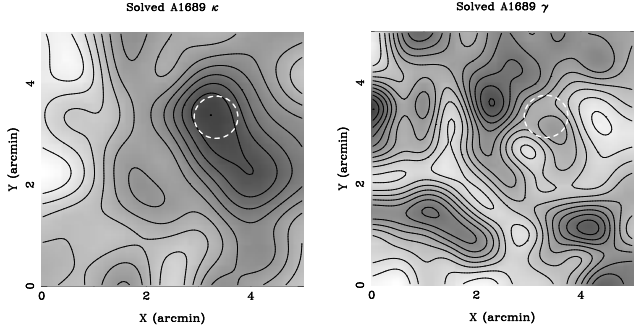}
\hfill}
\epsfverbosetrue
\small\caption{A1689 solved convergence and shear distributions. Darker areas
represent a higher distribution density. White dashes show the observed
critical line. The plots are smoothed from a 12 by 12 grid with north up
and east to the left.}
\label{fig3}
\end{figure}

Figure \ref{fig3} shows the solved mass density
and shear distribution. Comparison with the mass density map
illustrated in T98 (figure 6) which was produced with the sheet 
$\kappa$ estimator shows very similar structure. We find that the 
value of $\kappa$ at the
peak calculated here is approximately 10\% lower than the peak value
in T98 since 
the sheet estimator over-estimates $\kappa$ inside
critical line regions. 
This has little effect on the total integrated mass of A1689 found in T98.
The $\gamma$ distribution
is shown for completeness although undoubtedly suffers from boundary
effects typically found in the models.

\section{Shear Analysis}
\label{sec-shear}

Having shown that pixellisation allows us to accurately reconstruct 
surface mass densities from magnification data, we now apply it to 
shear analysis.
Shear analysis exploits the idea that a given distribution of images of
galaxies lying behind a lensing cluster will, in the statistical mean,
have regions of lens-induced correlations in image orientation and
ellipticity. Measuring the quadrupole moments of individual galaxy
images enables the construction of a map of the ellipticity
parameters, $e_{ij}$ (Valdes, Tyson \& Jarvis 1983).
The ellipticity parameters
relate to the surface mass density and shear via
(Kaiser 1995, K95 hereafter):
\be
\label{ellip_pars}
e_{ij}=\frac{\gamma_{ij}}{1-\kappa} , \quad
\gamma_{ij}=\left( \begin{array}{cc} \gamma_{1} & \gamma_{2} \\
\gamma_{2} & -\gamma_{1} \\ \end{array} \right)
\ee
One way of solving this for $\kappa$ in the weak lensing regime
is to follow the approach of Kaiser \& Squires (1993).
Generalisations of this to the strong regime have been 
made by K95. One would have hoped that an
alternative to such approaches would be to pixellise equation 
(\ref{ellip_pars}) and use equation (\ref{pix_gam}) to solve it by 
matrix inversion. However,
the resulting matrix equation is ill-conditioned,
since the matrix $D^{mn}_{1}$ is singular and $D^{mn}_2$ is itself
ill-conditioned.
Instead, we show a new, simplified expression for the solution to 
Kaisers' ellipticity equation and then pixellise it.

Starting with the equation (K95),
\be
\partial_{i}\kappa = \partial_{j}\gamma_{ij}
\ee
and using equation (\ref{ellip_pars}), one can show that 
\be
\label{log_kap}
\partial_{i}\ln(1-\kappa)=-\partial_{j}\ln(\delta_{ij}+e_{ij}).
\ee
The term on the right hand side is obtained from the definition,
\be
\ln(I+\bB)=\bB+\frac{1}{2}\bB^2+\frac{1}{3}\bB^3+\cdots
\ee
where $I$ is the identity matrix and $\bB$ is an arbitrary square matrix.
Using this expansion and collecting even and odd terms we find,
\be
\label{eqseries}
 \ln(\delta_{ij}+e_{ij}) = -\frac{1}{2} \ln(1-e^2) \delta_{ij} + \frac{1}{2}
	\ln \left(\frac{1+e}{1-e}\right) \frac{e_{ij}}{e},
\ee
where $e^2=e_1^2+e_2^2$ and $e_{i}=\gamma_{i}/(1-\kappa)$.
This result requires that $e<1$.
Inserting equation (\ref{ellip_pars}) into the magnification 
equation (\ref{mageq}) we find
\be
	A^{-1} = |(1-\kappa)^2(1-e^2)|.
\ee
Hence the parity changes when $e>1$. Since  
$e_{ij}$ and $e^{-1}_{ij}$ are observationally indistinguishable and 
flip from one to another whenever there is a parity change, we
can satisfy the criterion $e<1$ just by noting the critical line positions
and inverting the ellipticity matrix when one is crossed.

Finally, inserting equation (\ref{eqseries}) into equation (\ref{log_kap}),
and solving for $\kappa$ we find the pixellised solution is 
\be
\label{kap_ellip}
\kappa_{n} = 1-(1-e_n^2)^{1/2}
\exp\left[-\frac{1}{2}(D^{nm}_{1}s^m_1+D^{nm}_{2}s_2^m)\right]
\ee
where
\be
s_i=\frac{e_{i}}{e}\ln\left(\frac{1+e}{1-e}\right),
\quad i=1,2
\ee

\begin{figure}
\epsfxsize=8.4cm
{\hfill
\epsfbox{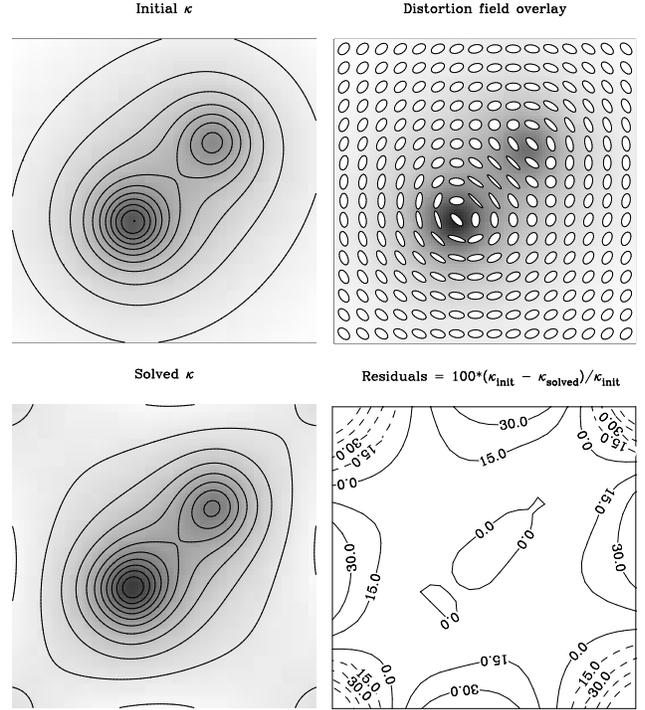}
\hfill}
\epsfverbosetrue
\small\caption{Reconstruction of $\kappa$ from the ellipticity parameters.
Contours are at the same levels in both $\kappa$ plots. The distortion field
is illustrated by plotting the apparent shape of an intrinsically circular
background object.}
\label{fig4}
\end{figure}

Equation (\ref{kap_ellip}) is the second main result of this letter. We
can directly calculate $\kappa$ given a measured ellipticity field. 
Figure \ref{fig4} shows the results of reconstructing $\kappa$ using
equation (\ref{kap_ellip}) for the dumb-bell model. The ellipticity 
parameters are calculated from equation (\ref{ellip_pars}) using the
$\kappa$ and $\gamma$ distribution. We normalise the reconstructed
$\kappa$ to both peaks in the initial $\kappa$ distribution. The residuals,
again being dominated by boundary effects,
show that reconstruction is possible to within approximately 10 percent across
the field of view.

\section{Summary}

We have outlined a method for directly calculating accurate,
self-consistent surface mass density and
shear distributions from the lens amplification and critical
line positions. The method has been demonstrated
with the isothermal sphere and dumb-bell cluster models.
We find it reconstructs the surface density to within a percent over 
most of the field of view. The reconstruction of the shear pattern
only has fractional accuracy of a few tenths due to boundary effects. 
We have applied the method to magnification data from Abell 1689, and 
reconstructed its surface mass and shear distribution.

We have also found a simplified solution to the problem of estimating
surface mass density from galaxy ellipticities. This approach puts the
calculation of surface mass from shear and magnification on
an equal footing, and we shall investigate the combined analysis
elsewhere.

\bigskip
\noindent{\bf ACKNOWLEDGEMENTS}
\bib\strut

\noindent
SD is supported by a PPARC studentship. ANT is a PPARC research associate. 
Thanks to our collaborators Tom Broadhurst, Txitxo Benitez, Eelco van 
Kampen for allowing us to use data from Abell 1689. We also thank 
Hanadi AbdelSalam for bringing the benefits of pixellisation to our attention
and for useful discussion.

\bigskip
\noindent{\bf REFERENCES}
\bib \strut

\bib AbdelSalam H.M., Saha P., Williams L.L.R., 1998, MNRAS, 294, 734

\bib Bartelmann M., Narayan R., Seitz S., Schneider P., 1996, ApJ, 464, 115

\bib Bartelmann M., Weiss A., 1994, A\&A, 287, 1

\bib Bridle S.L., Hobson M.P., Lasenby A.N., Saunders R., 1998, 
astro-ph/9802159

\bib Broadhurst T.J., Taylor A.N., Peacock J.A., 1995, ApJ, 438, 49

\bib Kaiser N., (K95), 1995, ApJ, 439, L1

\bib Kaiser N., Squires G., 1993, ApJ, 404, 441

\bib Seitz C., Schneider P., 1995, A\&A, 297, 287

\bib Seitz C., Schneider P., Bartelmann M., 1998, astro-ph/9803038

\bib Squires G., Kaiser N., 1996, ApJ, 473, 85

\bib Taylor A.N., Dye S., Broadhurst T.J., Benitez N., van Kampen E., (T98),
     1998, ApJ, 501, 539

\bib Tyson J.A., Valdes F., Wenk R.A., 1990, ApJ, 349, L1

\bib Valdes F., Tyson J., Jarvis J., 1983, ApJ, 271, 431

\bib van Kampen, 1998, submitted to MNRAS

\end{document}